\documentclass[12pt,preprint]{aastex63}
\usepackage{rotating}
\usepackage{epsfig}

\begin{document}

\title{A Massive Young Runaway Star in W49 North}

\author[0000-0003-2737-5681]{Luis F. Rodr{\'\i}guez}
\affil{Instituto de Radioastronom\'\i a y Astrof\'\i sica, 
Universidad Nacional Aut\'onoma de M\'exico,, Apdo. Postal 3-72 (Xangari), 58089 Morelia, Michoac\'an, M\'exico.}
\affil{Mesoamerican Center for Theoretical Physics, Universidad
Aut\'onoma de Chiapas, Carretera Emiliano Zapata Km.4,
Real del Bosque (Ter\'an). 29050 Tuxtla Guti\'errez, Chiapas,
M\'exico}

\author[0000-0003-1480-4643]{Roberto Galv\'an-Madrid}
\affil{Instituto de Radioastronom\'\i a y Astrof\'\i sica, 
Universidad Nacional Aut\'onoma de M\'exico,, Apdo. Postal 3-72 (Xangari), 58089 Morelia, Michoac\'an, M\'exico.}

\author{Joel Sanchez-Bermudez}
\affil{Instituto de Astronom\'ia, Universidad Nacional Aut\'onoma de M\'exico, Apdo. Postal 70264, Ciudad de M\'exico 04510, M\'exico.}
\affil{ Max-Planck-Institut für Astronomie, Königstuhl 17, 69117 Heidelberg, Germany.}

\author[0000-0003-3115-9359]{Christopher G. De Pree}
\affil{Department of Physics \& Astronomy, Agnes Scott College, 141 East College Avenue, Decatur, GA 30030, USA.}

\email{l.rodriguez@irya.unam.mx}
 
\begin{abstract}

We analyzed high angular resolution 45.5 GHz images of the W49 North massive star forming region obtained in 1998 and 2016 with the Very Large Array.
Most of the ultracompact HII regions show no detectable changes over the time interval of the observations. However, subcomponents
B1, B2, G2a and G2c have increased its peak flux densities by values in the range of 3.8 to 21.4 \%. Most interestingly, the cometary
region C clearly shows proper motions that at the distance of the region are equivalent to a velocity of 76$\pm$6 km s$^{-1}$ in the plane of
the sky. We interpret this region as the ionized bowshock produced by a runaway O6 ZAMS star that was ejected from the eastern edge of Welch's ring 
about 6,400 years ago.

\end{abstract}  


\section{Introduction}

Dynamical interactions in young stellar systems have important consequences to cluster- and stellar evolution \citep{ZY07,Reipurth00,Tan2014}. An important signature of such interactions is the existence of a population of runaway stars \citep{Poveda67}. 
A significant fraction of the field O-type stars within a few kpc of the Sun could be runaways \citep{MaizApellaniz04}. A few massive stars have been reported as runaways in optically-visible clusters such as R136 in the Large Magellanic Cloud \citep{Lennon18} and Cygnus OB2 in our Galaxy \citep{Comeron07}, inferring travel times $\gtrsim 1$ Myr, or about the cluster age, which suggests that dynamical ejections are simultaneous to cluster formation. Indeed, for the case of younger, more obscured regions, \citet{Rodriguez05} reported the dynamical ejection of three massive (proto)stars seen in the radio continuum in the Orion BN/KL region, with a travel time $\sim 500$ yr. Refined measurements confirmed the original findings \citep{Rodriguez17} and unveiled additional runaway candidates with 
similar ejection times \citep{Dzib17,Luhman17}. Other young runaway stars  in the vicinity of the Orion Nebula Cluster with longer ejection times have been reported by \citet{McBrideKounkel19} and \citet{Dzib17}.

In this Letter, we report on the existence of a massive runaway moving away from the Welch ring of Ultracompact (UC) HII regions \citep{Welch87,DePree97} in W49N, one of the most massive, concentrated, and luminous cluster forming regions in the Galaxy \citep{GM13,Lin16}. We also report on the flux variability of the 7 mm continuum in several of the UC HIIs in the field, and discuss them in the context of the variability study at 3.5 cm presented in \citet{DePree18}. 

\section{Multiepoch Data}

We analyzed high-angular resolution Q-band VLA data. 
The older image comes from concatenating data of three projects  (AD356, 1995 May 30, D-configuration; AD414, 1998 Apr 11, A-configuration; and TEST, 2001 Apr 09, B-configuration). Since most of the data comes from observations in the A configuration we attribute to these concatenated data an epoch of 1998.28. The more recent Q band image was taken as part of Project 16B-022 on 2016 Sep 28 (epoch = 2016.74) with the VLA in the A configuration. The gain calibrator for 1998 Apr 11 and 2016 Sep 28 was the same, J1922+1530. However, the positions for the gain calibrator between these two epochs differed by $\sim0\rlap.{''}05$. As we will see below, the method used to align the two epochs will correct for these small differences, as well as others coming from tropospheric phase noise and self-calibration.

Both images were taken at a central frequency of  45.5 GHz, with a total bandwidth of 100 MHz for the 1998.28 data and of  2048 MHz for the 2016.74 data. The 1998.28 data were calibrated using the
standard AIPS (Astronomical Image Processing System) techniques, while the 2016.74 data were calibrated using the CASA (Common Astronomy Software Applications) package of NRAO and
the pipeline provided for VLA\footnote{https://science.nrao.edu/facilities/vla/data-processing/pipeline} observations. The data were self-calibrated and the resulting images have angular resolutions of $0\rlap.{''}055 \times 0\rlap.{''}044$; 
PA = $8\rlap.^\circ6$ and $0\rlap.{''}081 \times 0\rlap.{''}041$; PA = $85\rlap.^\circ5$, for 1998.28 and 2016.74, respectively. Both images were smoothed to an angular resolution of $0\rlap.{''}1 \times 0\rlap.{''}1$ to allow a more accurate comparison between them. We also removed in both images faint extended emission with scales larger that $\sim1{''}$ using the AIPS task MWFLT and following the prescription of \citet{Rudnick02}.

\section{Image Analysis}

Comparison of the two images showed an offset of the order of $0\rlap.{''}1$ between them. This cannot be attributed to proper motions of the region since these motions have a magnitude of $\sim$6 mas yr$^{-1}$ 
\citep{Zhang13} and over the 18 year interval between the images it will accumulate to a displacement of order $0\rlap.{''}01$.  Most likely the displacement between images is due to a combination of the tropospheric conditions during the observations, the small position displacements that can result from self-calibration and the difference in position for the gain calibrator in the two epochs. To align the two images as best as posible we followed two procedures. The positions of the most compact and bright sources (components A, B2 and G2a in the nomenclature of \citet{Dreher84} and \citet{DePree97}) were compared, finding that the 1998 image was displaced by $\Delta \alpha = -0\rlap.{''}031 \pm 0\rlap.{''}004$; $\Delta \delta = -0\rlap.{''}118 \pm 0\rlap.{''}009$. The second procedure consisted of subtracting the 1998.28 image from the 2016.74 image adding displacements in increments of $0\rlap.{''}01$ both in $\alpha$ and $\delta$ and searching for the residual image with the smallest rms. A displacement of $\Delta \alpha = -0\rlap.{''}02$; $\Delta \delta = -0\rlap.{''}11$ was found to be optimal  and we adopted it to compare the two images.

\begin{figure}
\centering
\vspace{-2.8cm}
\includegraphics[angle=0,scale=0.45]{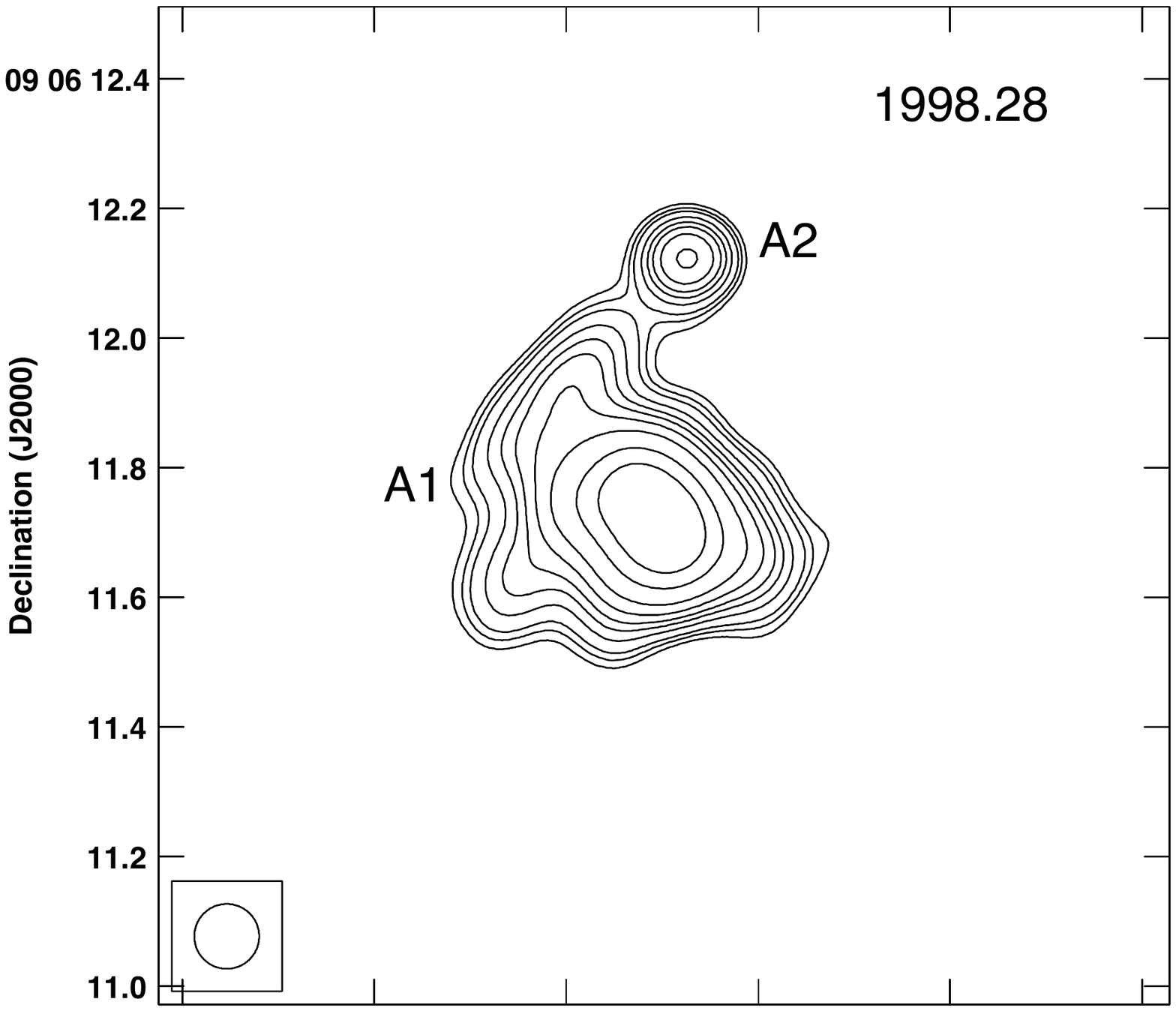}
\vskip-5.8cm
\includegraphics[angle=0,scale=0.45]{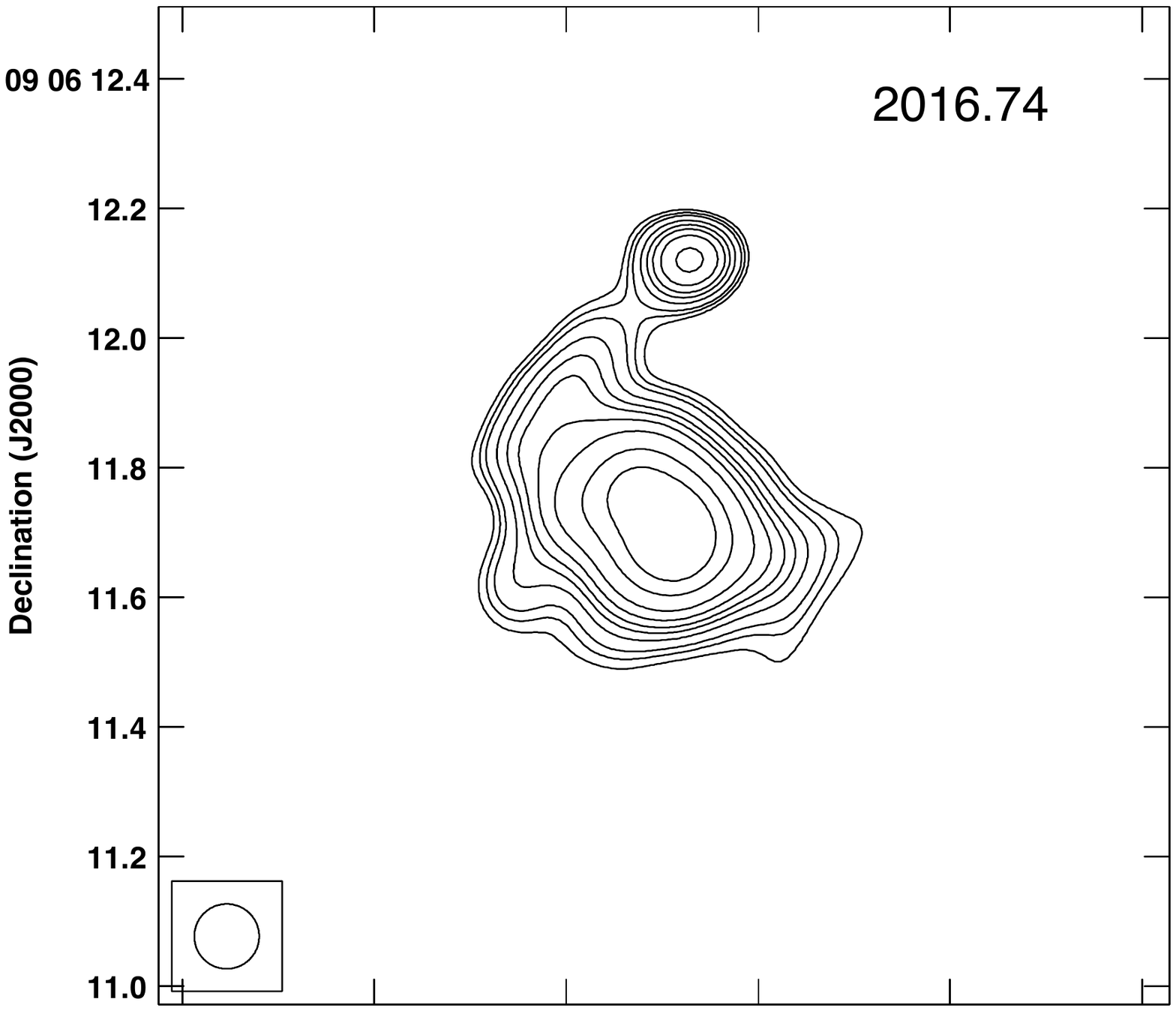}
\vskip-5.8cm
\includegraphics[angle=0,scale=0.45]{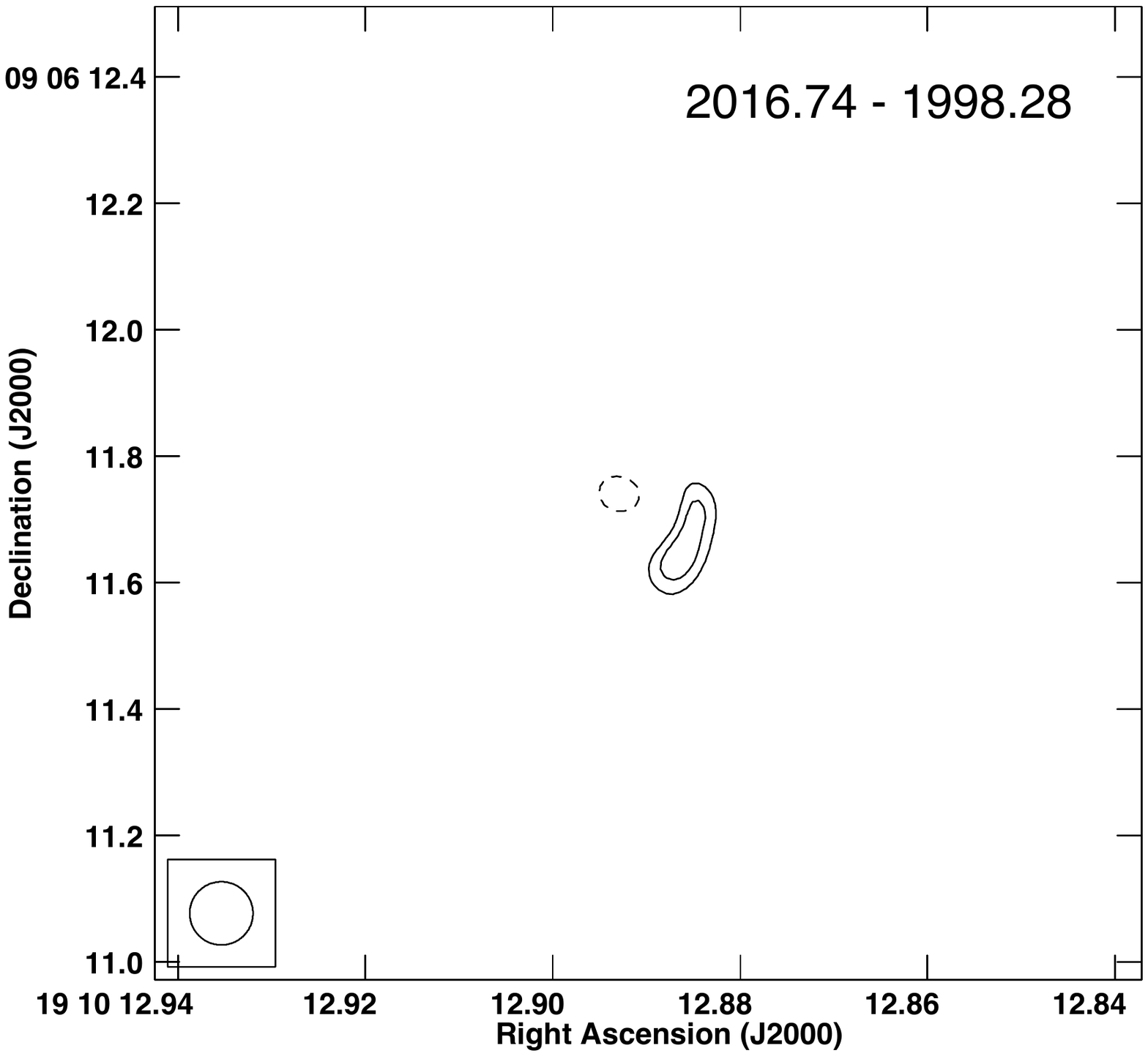}
\vskip-2.5cm
\caption{\small Images of the components A1 and A2 for 1998.28 (top), 2016.74 (middle) and difference image (bottom). The beam ($0\rlap.{''}1 \times 0\rlap.{''}1$) is shown in the bottom left corner.
The contours are -4, 4, 5, 6, 8, 10, 12, 15, 20, 30, and 40 times 1.6, 1.6 and 1.8 mJy beam$^{-1}$, respectively.. The residuals in the difference image are not considered significant.
}
\label{fig1}
\end{figure}

Finally, the absolute flux density calibration at Q band is known to present random variations of the order of 10\%. Again we searched for the image subtraction that produced the smallest rms and this was obtained multiplying the 1998.28 image by a correction factor of 1.09. The final difference image shows significant residuals only in association with five components: B1, B2, C, G2a and G2c. Bright sources such as components A1 and A2 cancel out in the subtraction of the two images
(see Figure \ref{fig1}), indicating that they have not experienced significant changes in position, morphology or emission.

\section{Flux Variability}

\begin{figure}
\centering
\vspace{-2.5cm}
\includegraphics[angle=0,scale=0.27]{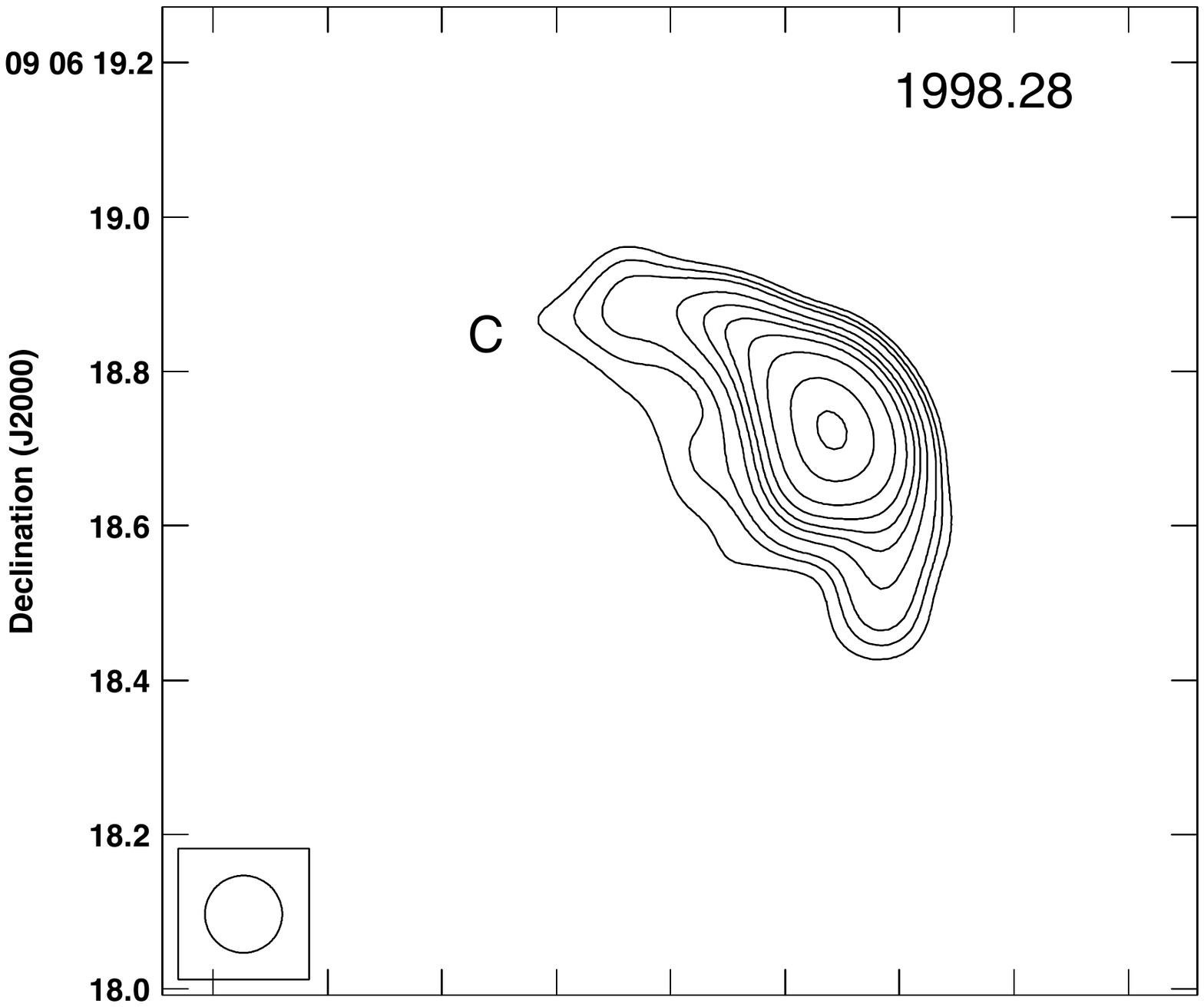}
\includegraphics[trim=3.2cm -0.6cm 0 0,angle=0,scale=0.26]{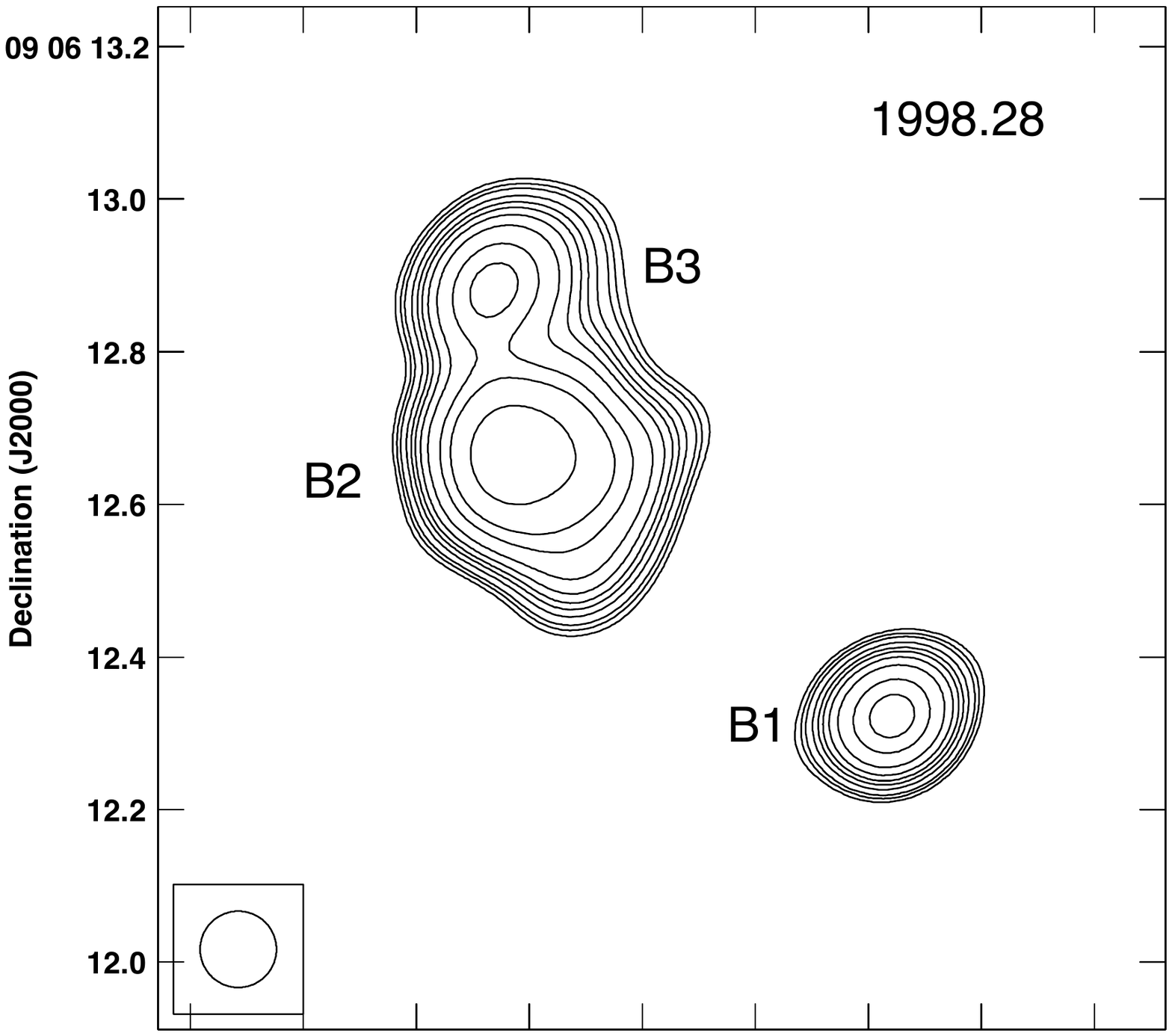}
\includegraphics[trim=3.2cm 2.2cm 0 0,angle=0,scale=0.32]{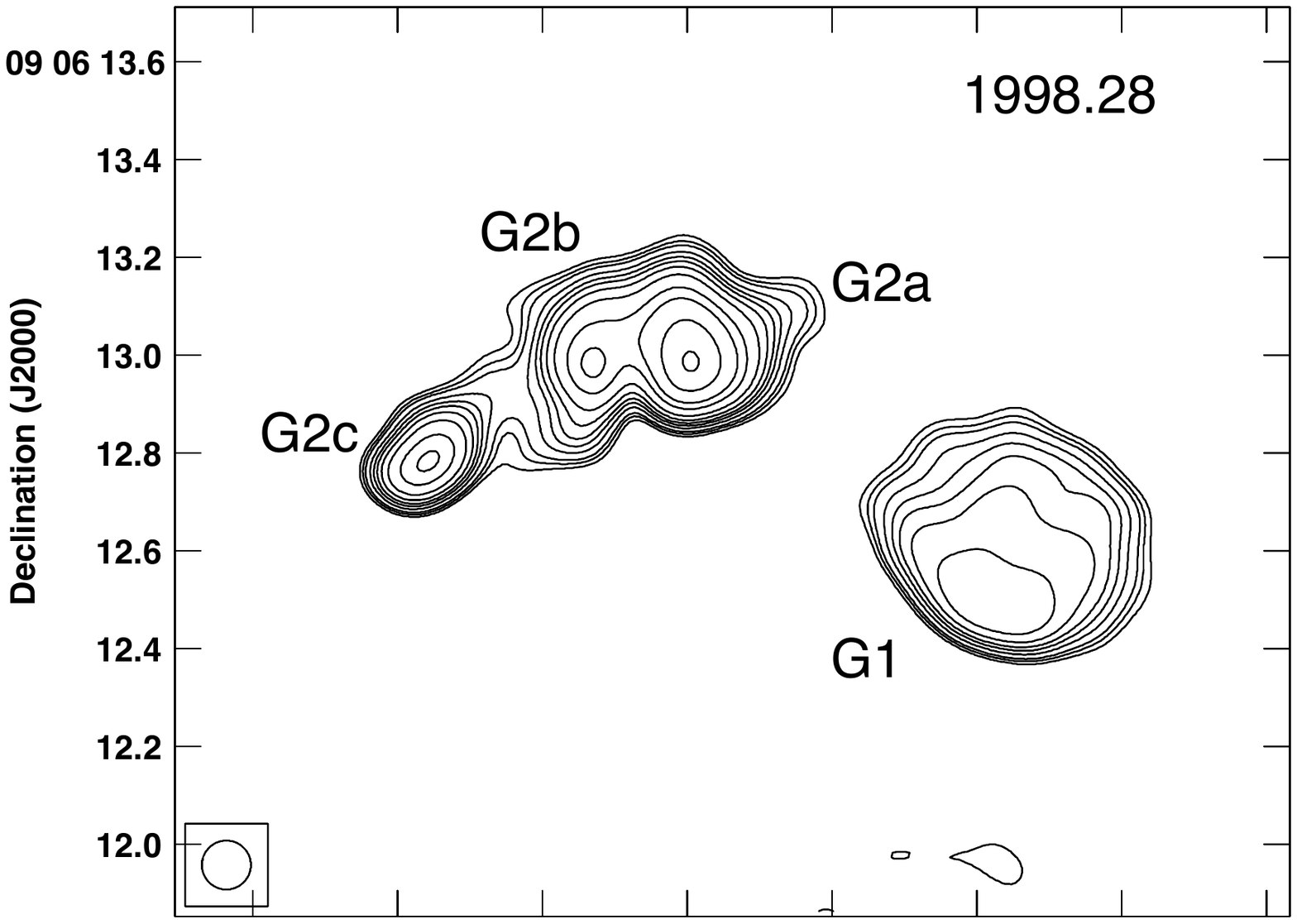}
\vskip-4.2cm
\includegraphics[angle=0,scale=0.27]{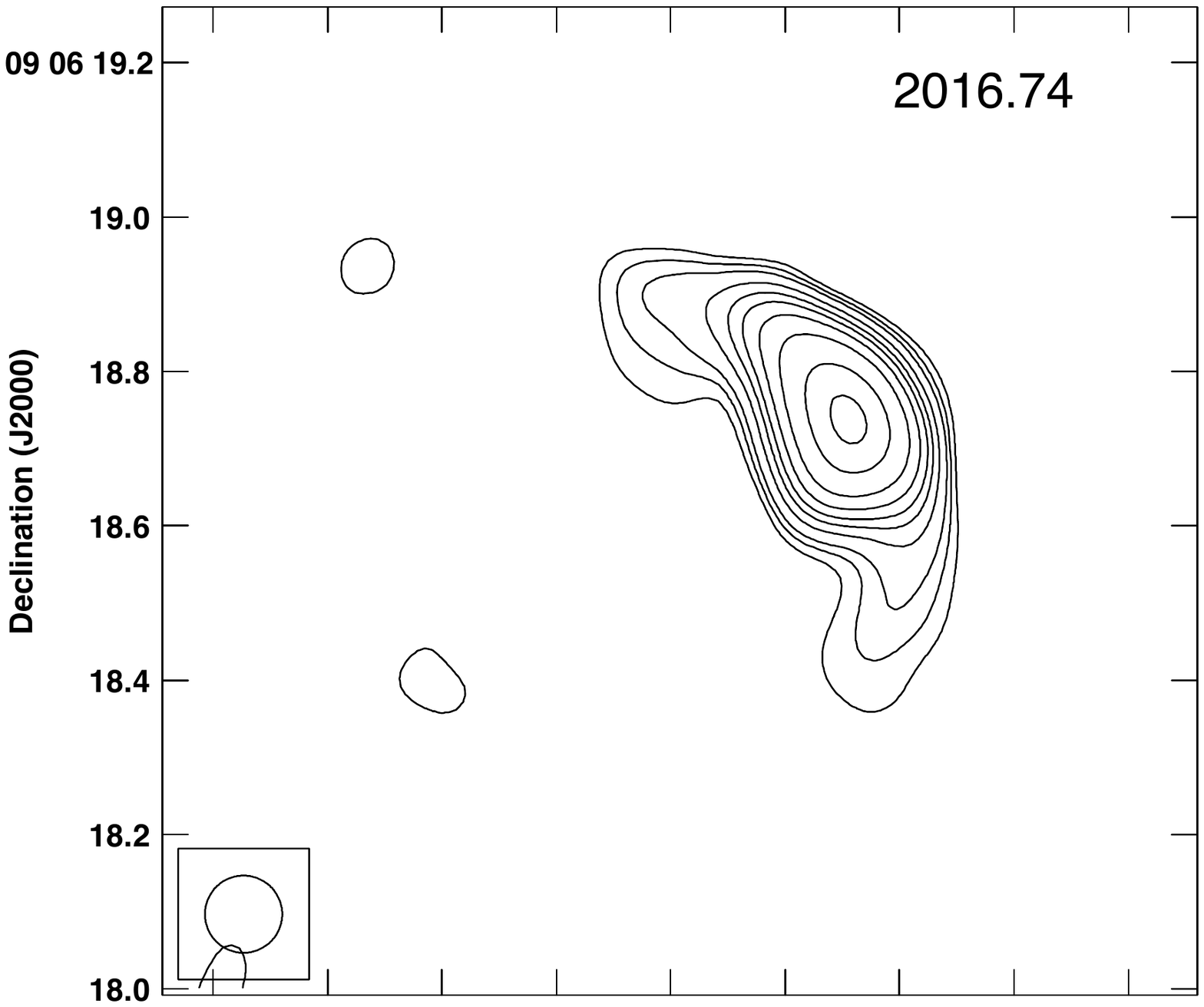}
\includegraphics[trim=3.2cm -0.3cm 0 0,angle=0,scale=0.26]{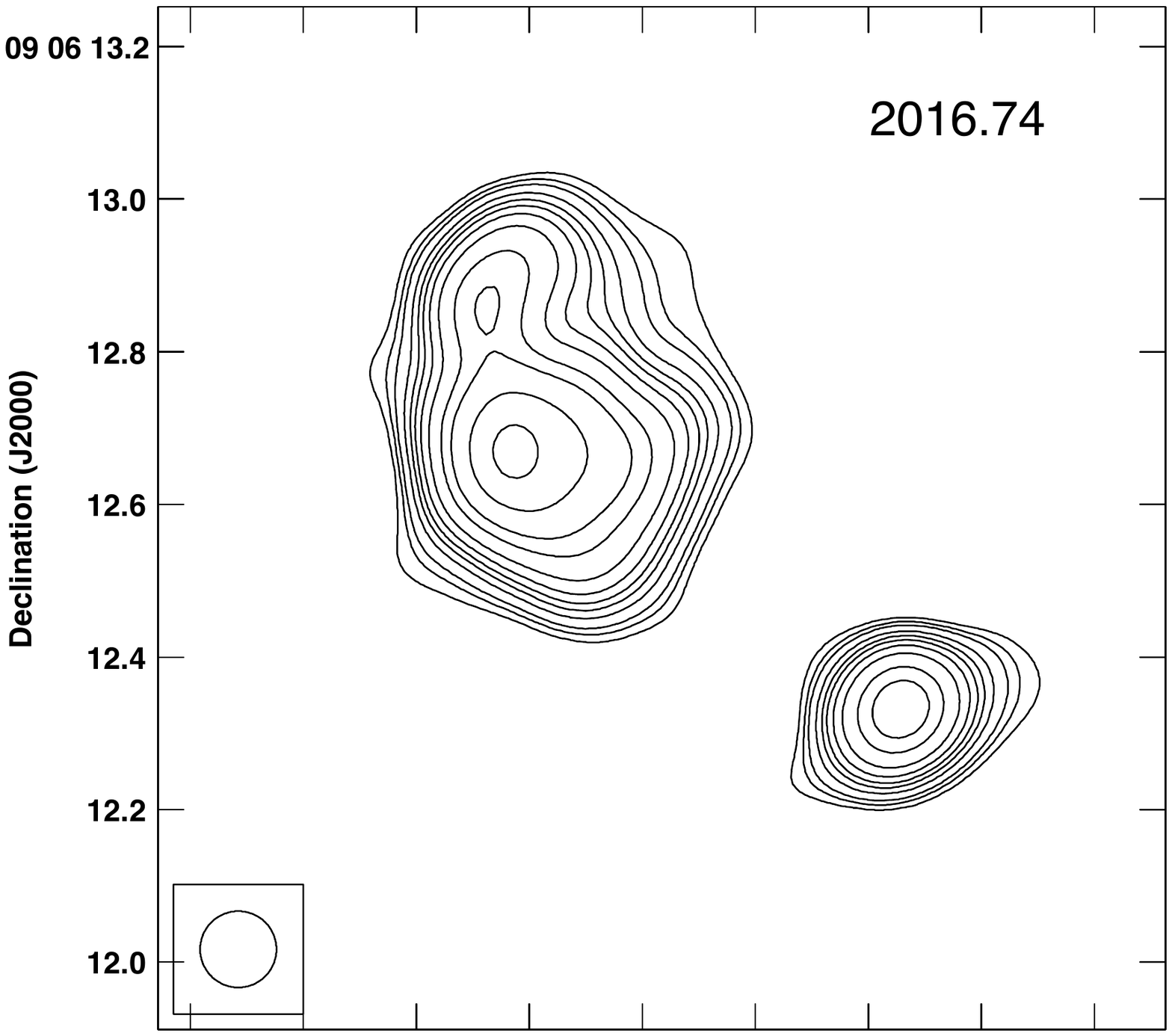}
\includegraphics[trim=3.2cm 1.8cm 0 0,angle=0,scale=0.32]{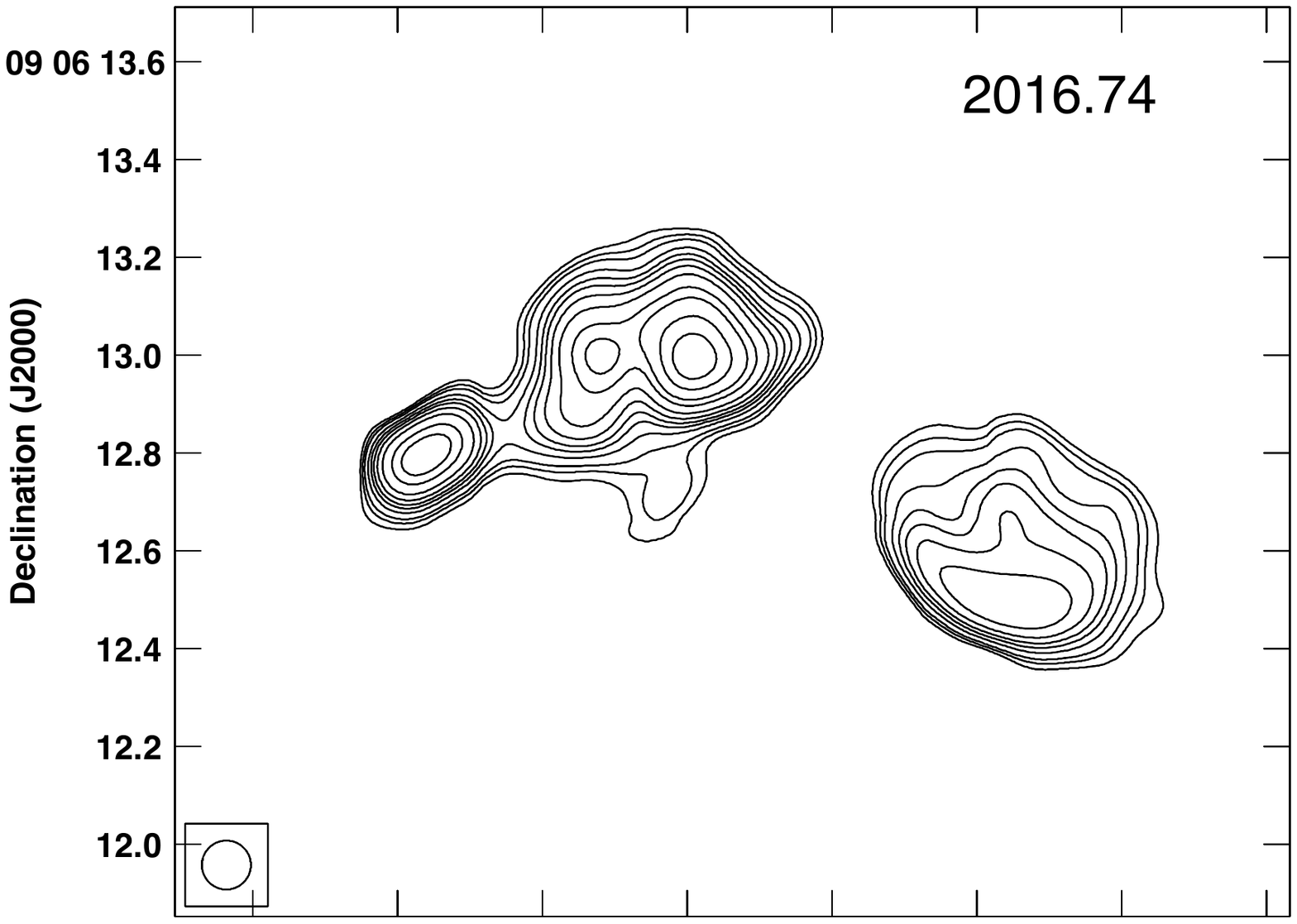}
\vskip-4.2cm
\includegraphics[angle=0,scale=0.27]{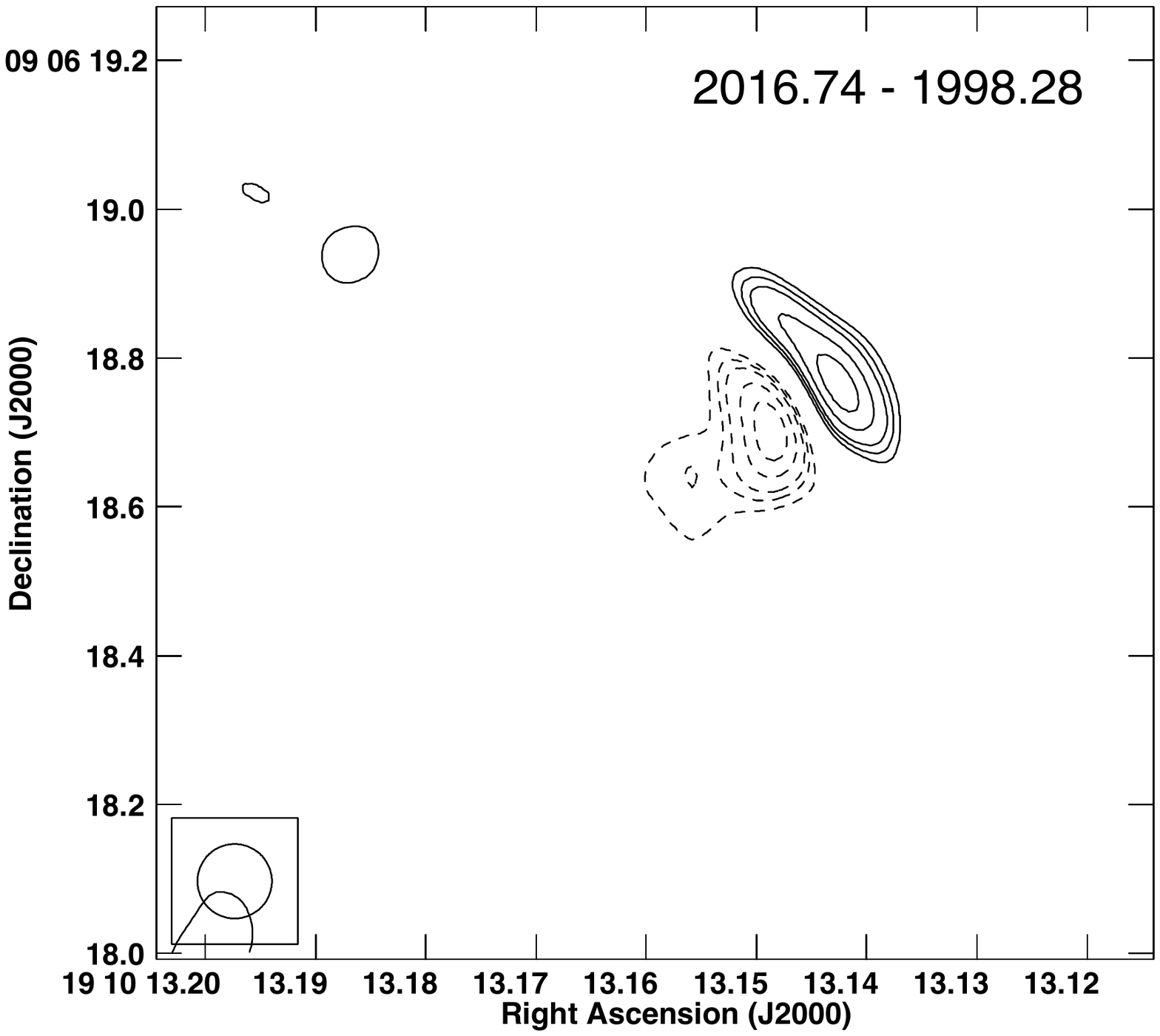}
\includegraphics[trim=3.2cm -0.3cm 0 0,angle=0,scale=0.26]{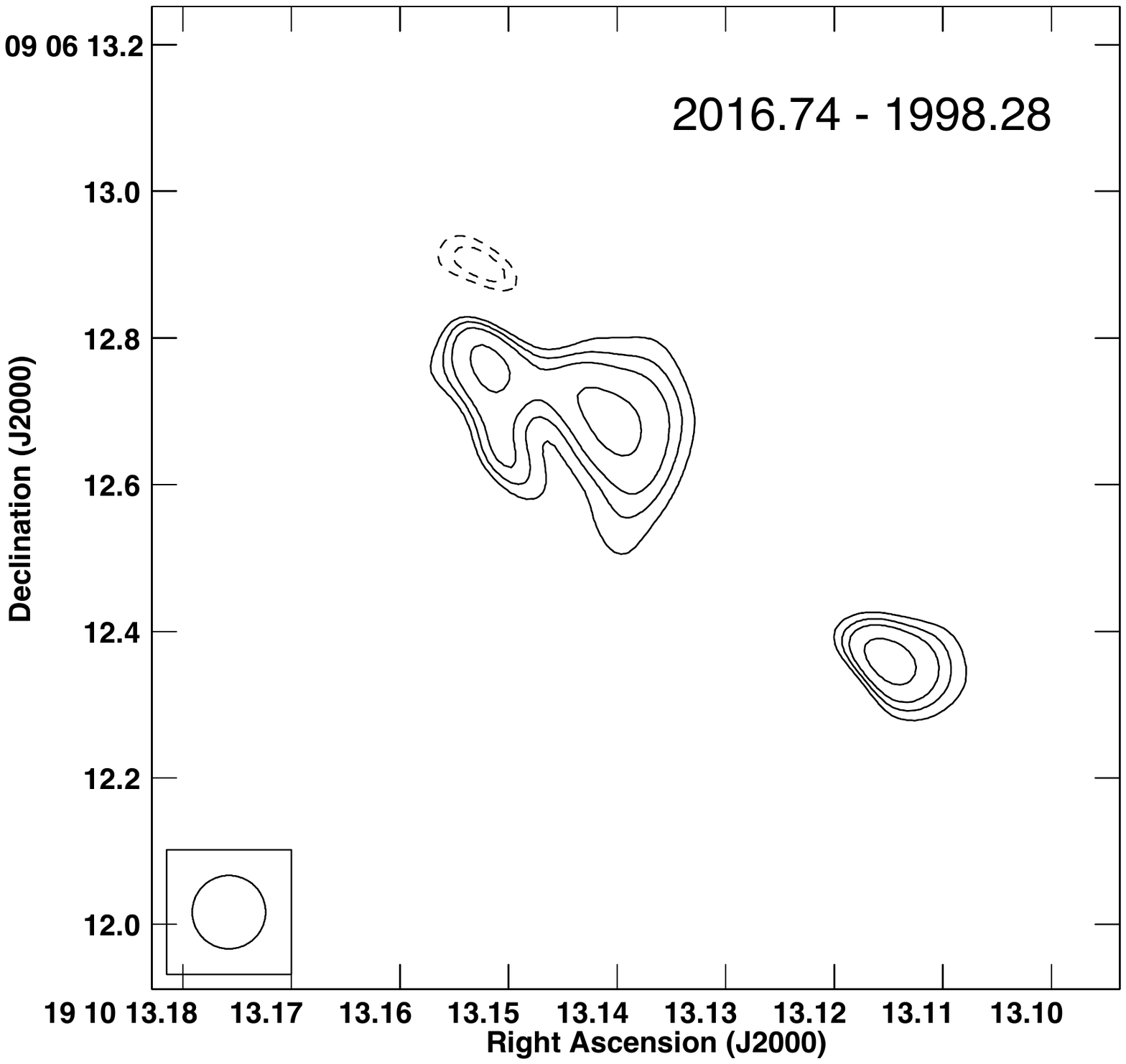}
\includegraphics[trim=3.2cm 1.8cm 0 0,angle=0,scale=0.32]{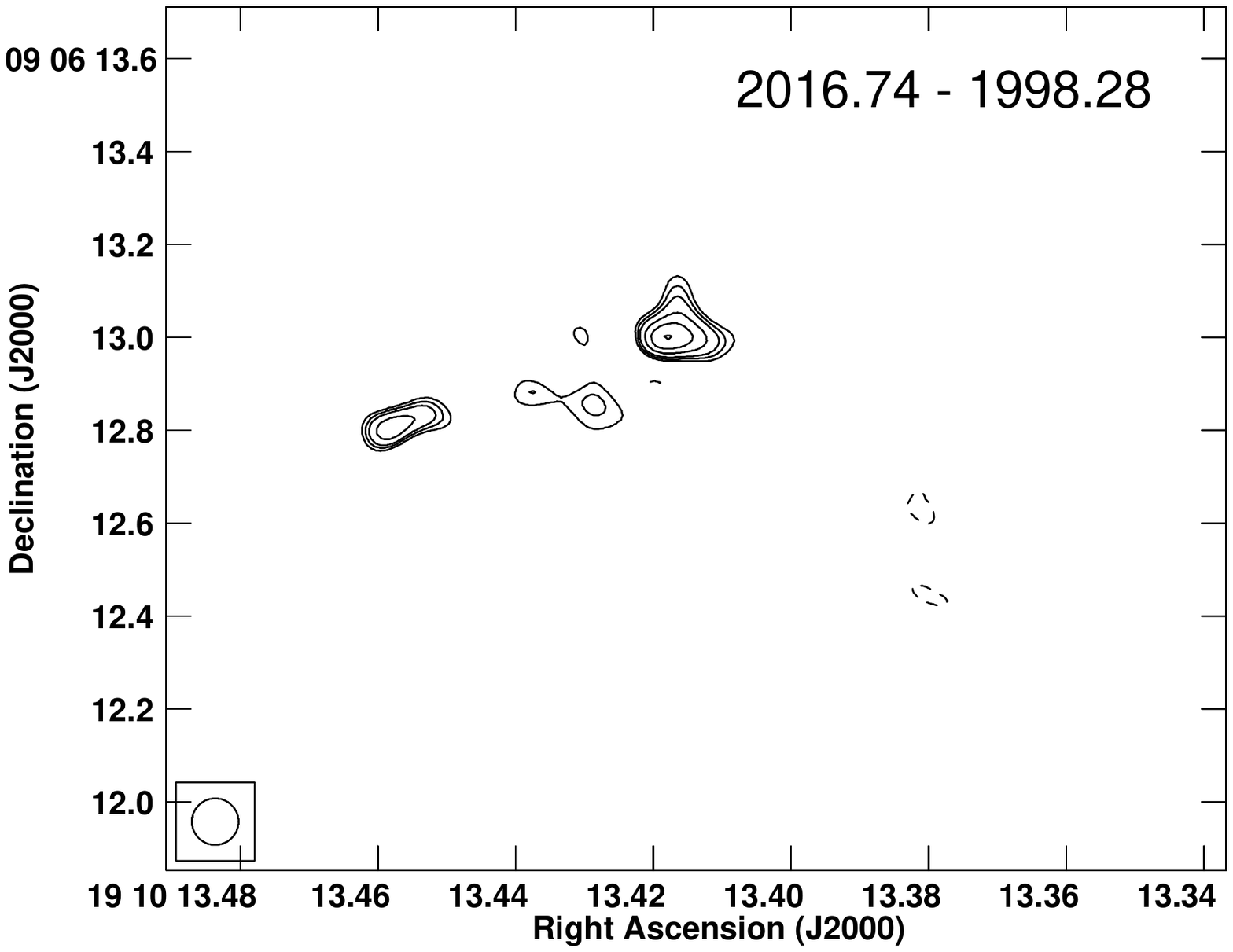}
\vskip-1.5cm
\caption{\small {\it Left column:} 
images of the component C for 1998.28 (top), 2016.74 (middle) and difference image (bottom).
The contours are -10, -8, -6, -5, -4, 4, 5, 6, 8, 10, 12, 15, 20, 30, and 40 times 1.6, 1.6 and 1.4 mJy beam$^{-1}$, respectively.  The residuals in the difference image are interpreted to imply
a motion of the source.
{\it Center column:} 
images of the components B1, B2 and B3 1998.28 (top), 2016.74 (middle) and difference image (bottom).
The contours are -5 -4, 4, 5, 6, 8, 10, 12, 15, 20, 30, 40 times and 60 times 1.6, 1.6 and 1.8 mJy beam$^{-1}$, respectively. The residuals in the difference image are interpreted to imply
an increase in the peak flux density of components B1 and B2.
{\it Right column:}
images of the component G1, G2a, G2b and G2c for 1998.28 (top), 2016.74 (middle) and difference image (bottom).
The contours are -4, 4, 5, 6, 8, 10, 12, 15, 20, 30, 40, 60 and 80 times 1.6, 1.6 and 2.2 mJy beam$^{-1}$, respectively. The residuals in the difference image are interpreted to imply
an increase in the peak flux density of components G2a and G2c.
The beam ($0\rlap.{''}1 \times 0\rlap.{''}1$) is shown in the bottom left corner of each panel. 
}
\label{fig2}
\end{figure}

The images associated with components C, B, and G are shown in Figure \ref{fig2}. 
In the case of the images of components B and G we find that the subcomponents B1, B2, G2a and G2c appear to have increased their peak intensity by 3.8$\pm$1.3\%, 4.8$\pm$0.7\%, 12.0$\pm$0.7\% and 21.4$\pm$1.4\%, respectively, over the time interval of the observations. Remarkably, over a similar time period (1994-2015), \citet{DePree18} found that at 3.6 cm component G decreased by 20\% in peak intensity. This result was obtained with an angular resolution
of $0\rlap.{''}8$ and it is difficult to directly compare with our results at $0\rlap.{''}1$ resolution. The difference images (bottom row of Fig. \ref{fig2}, center and right columns)
show that there are no measurable changes in morphology or position for the G and B sources (that would appear as a set of adjacent negative/positive contours), only the flux increment previously described.

A flux density increment at 7 mm accompanied by a decrement at 3.6 cm can be explained if the radio continuum is  -- partially -- optically thin at the shorter- and optically thick at the longer wavelengths. In this scenario, a sudden increase in the amount of molecular gas to be ionized could result in an ultracompact HII region that is smaller and denser \citep{Peters10,GM08}. 
Since the flux at optically thick wavelengths depends mainly on the angular size and in the optically thin regime depends more strongly on $n_e^2$, the small HII region could appear fainter at 3.6 cm and brighter at 7 mm. 
Altogether, the results of \citet{DePree18} and ours are consistent with the predictions of \citet{GM11}, who used the radiation-hydrodynamical simulations of \citet{Peters10} to estimate a 
$\sim$15\% probability for an increase- and a $\sim$6\% probability for a decrease in the 2-cm flux of UC/HC HII regions over a time span of 20 years. The number of sources with detected time variability is also roughly consistent with the estimations of \citet{GM11}, who concluded that $\sim$10\% of the sources should have detectable variations within a period of $\sim 10$ years. 

\section{Radio Source C as a Runaway from W49N}   

\begin{figure}
\centering
\vspace{-2.8cm}
\includegraphics[angle=0,scale=0.45]{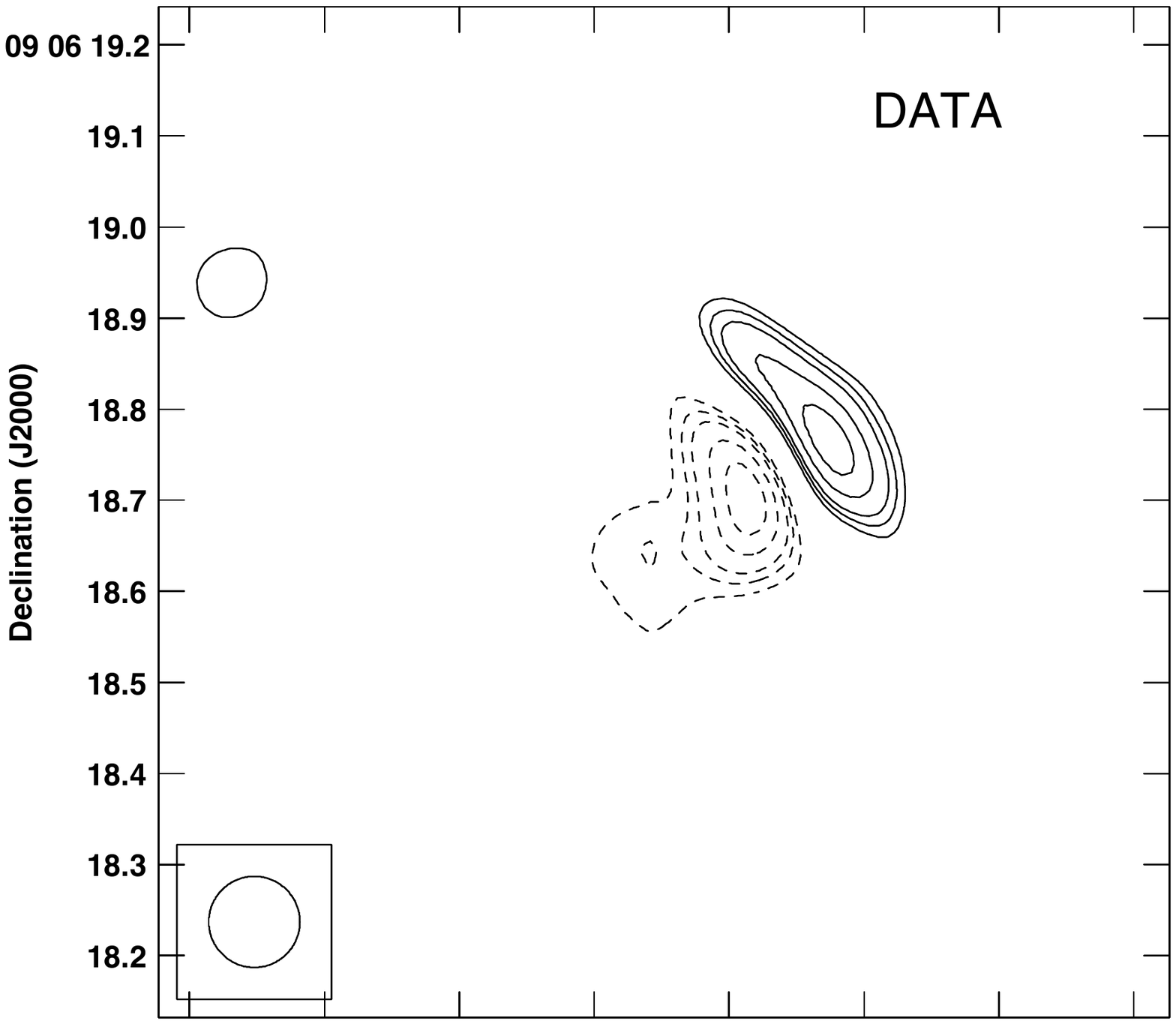}
\vskip-5.7cm
\includegraphics[angle=0,scale=0.45]{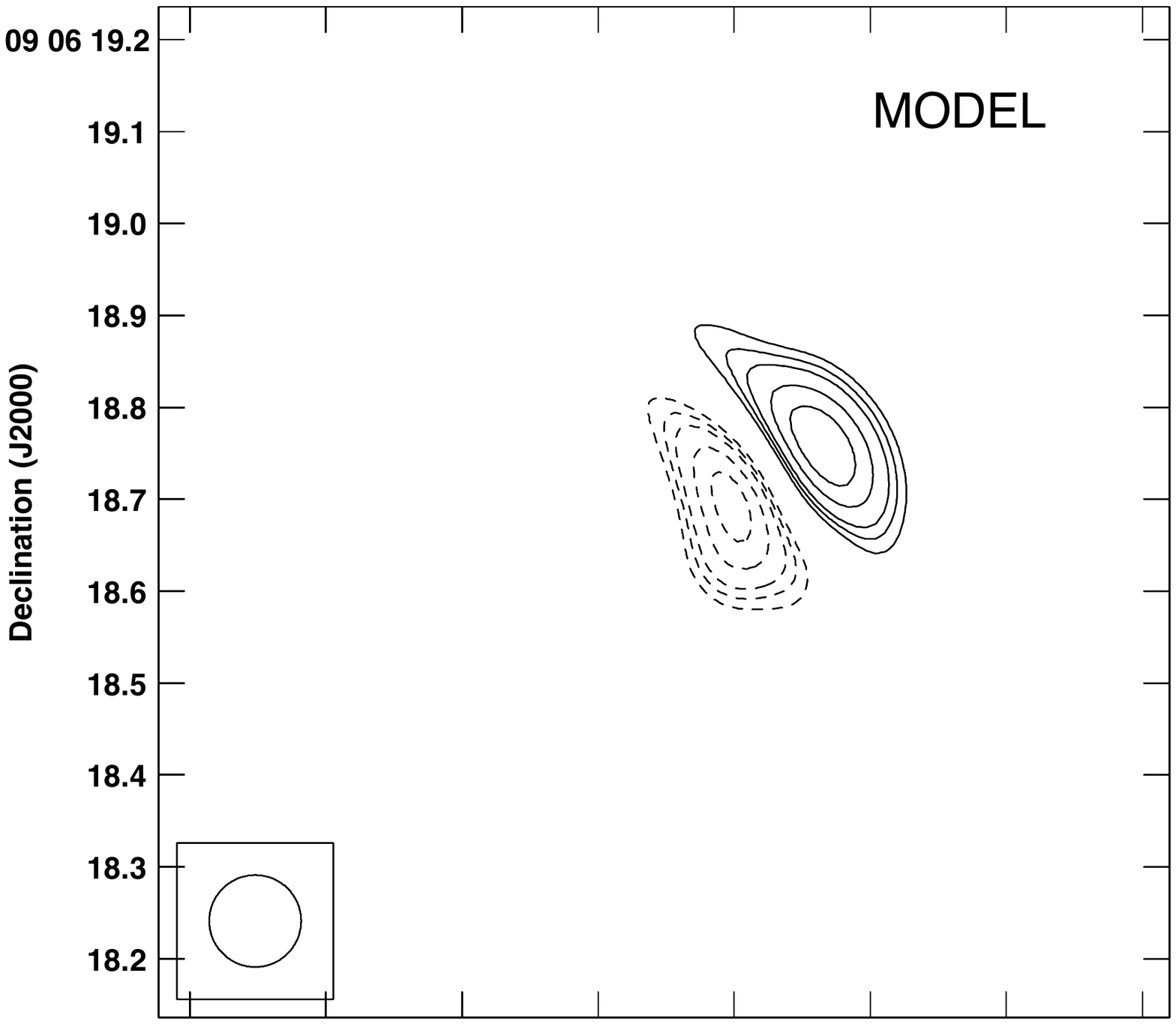}
\vskip-5.7cm
\includegraphics[angle=0,scale=0.45]{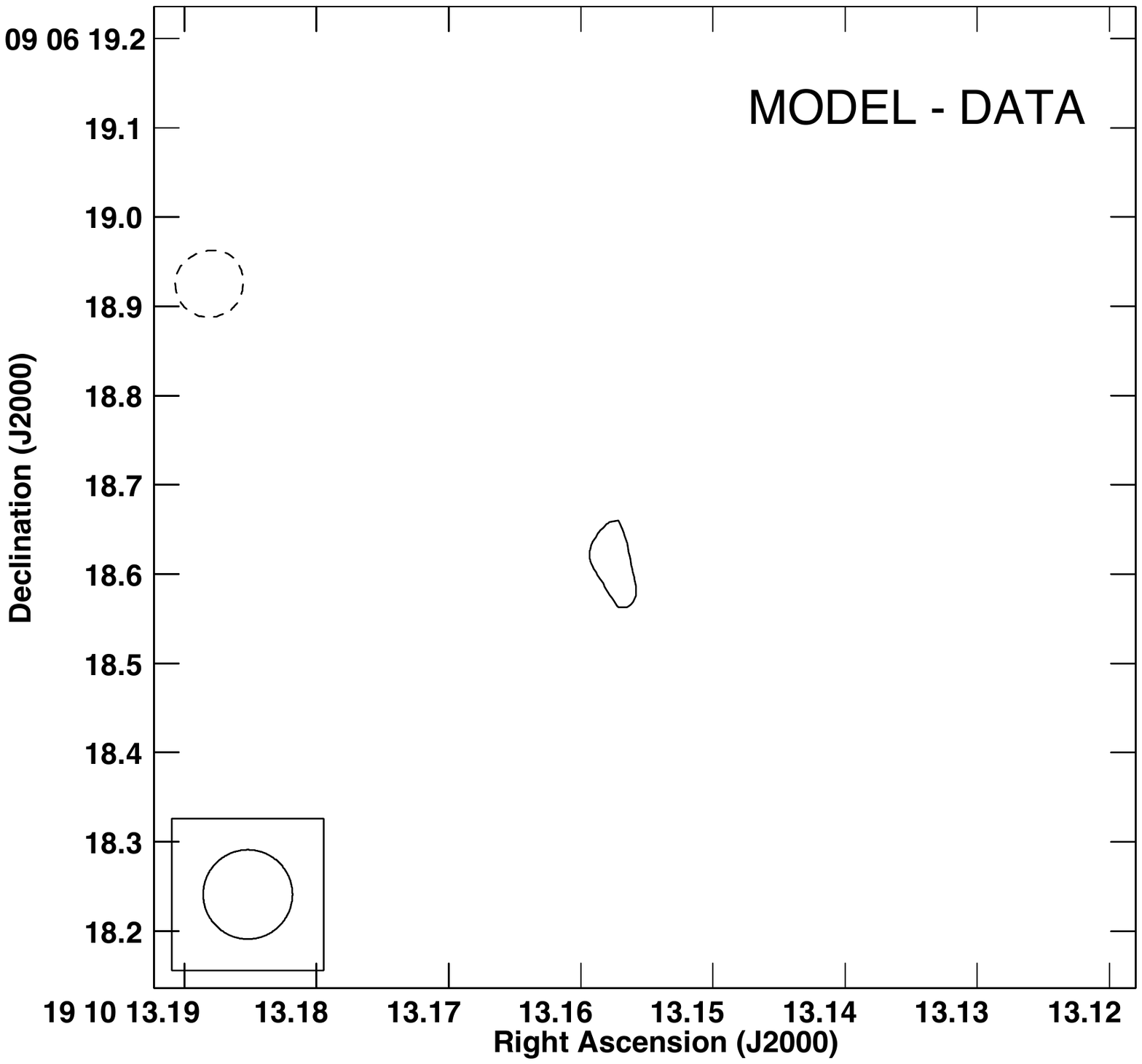}
\vskip-2.7cm
\caption{\small Image difference for the component C (top), model for the component C made as described in the text (middle) and
difference of the model minus data (bottom). The beam ($0\rlap.{''}1 \times 0\rlap.{''}1$) is shown in the bottom left corner.
The contours are -10, -8, -6, -5, -4, 4, 5, 6, 8, and 10 times 1.4 mJy beam$^{-1}$.
}
\label{fig3}
\end{figure}

In contrast to the components discussed above, component C clearly shows proper motions, signaled by the shifted negative and positive contours in Fig. \ref{fig3}, present at a 10-$\sigma$ significance  level. To estimate the angular size of the displacement, again we took the 1998.28 image and displaced it
by increments of $0\rlap.{''}001$ both in $\alpha$ and $\delta$. The initial image was subtracted to the displaced image until the residuals were minimized. The best agreement, shown in Fig. \ref{fig3}, was obtained with a
displacement of  $\Delta \alpha = -0\rlap.{''}020 \pm 0\rlap.{''}002$; $\Delta \delta = +0\rlap.{''}016 \pm 0\rlap.{''}002$. The total displacement is $\Delta s = 0\rlap.{''}026 \pm 0\rlap.{''}002$ at a position angle of $309^\circ \pm 4^\circ$. This
displacement is equivalent to a proper motion of 1.41$\pm$0.11 mas yr$^{-1}$. 

\begin{figure}
\centering
\vspace{-4.8cm}
\includegraphics[angle=0,scale=0.8]{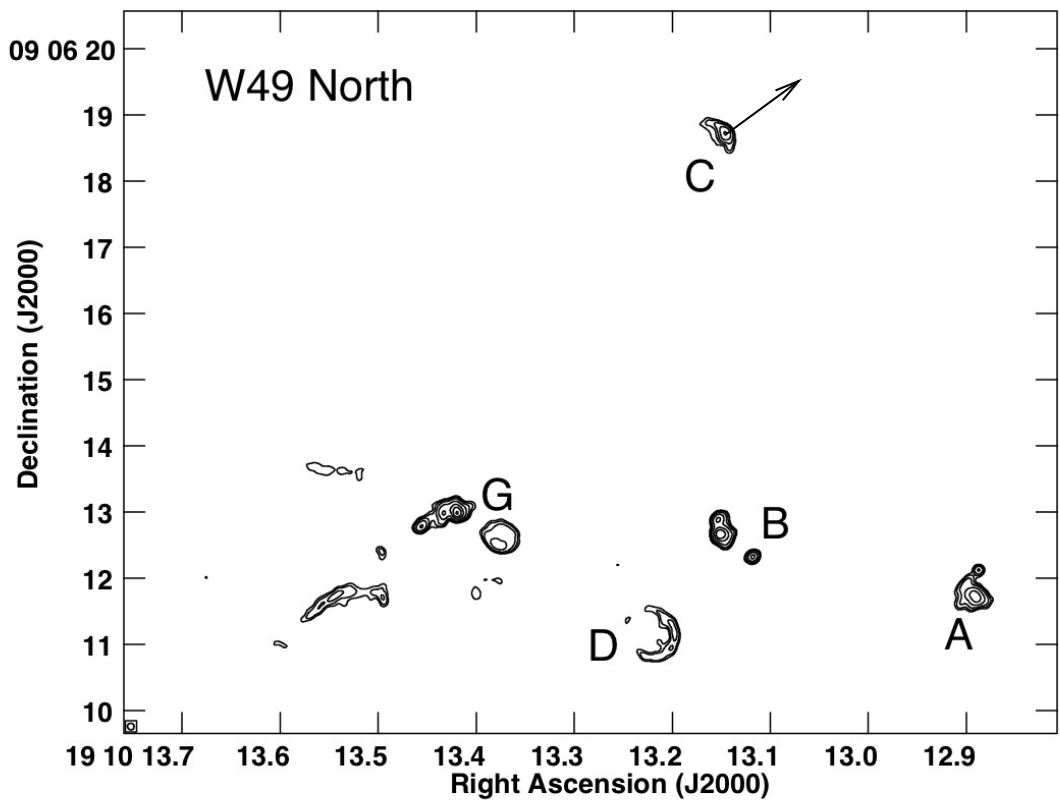}
\vskip-4.7cm
\caption{\small Image of W49 North for epoch 1998.28. The beam ($0\rlap.{''}1 \times 0\rlap.{''}1$) is shown in the bottom left corner.
The contours are 4, 6, 10, 20, 40, 60, 80 and 100 times 1.6 mJy beam$^{-1}$. The arrow indicates the displacement of source C for a time
span of $10^3$ years.
}
\label{fig4}
\end{figure}

The direction of the proper motion of source C matches the axis of its arc morphology (see Figs. \ref{fig2} and \ref{fig4}), suggesting that the emission arises from a bowshock \citep[e.g.,][]{ArthurHoare06}. 
At a distance of 11.1 kpc \citep{Zhang13}, the proper motions of component C over the time interval of 18.46 yr are equivalent to a velocity in the plane of the sky of 76$\pm$6 km s$^{-1}$. The position angle of the proper motions suggests that the star was ejected from the eastern edge of Welch's ring (see Figure \ref{fig4}). The present angular distance between the component C and the eastern edge of Welch's ring is $\sim 9{''}$, implying that this star was ejected about 6,400 years ago. 
The other sources in the region do not show significant evidence for proper motions, with an upper limit to their plane-of-the sky velocities of $\sim$20 km s$^{-1}$.

To further test the hypothesis of a bow shock from a runaway star, we modelled the emission profile using the algebraic solution of \citet{Canto96}. First, a radial profile of the emission was extracted at 120 different position angles. The position of the emission centroid and its uncertainty at each P.A. was obtained through Gaussian fitting. The derived bowshock profile consists of the locus containing all the fitted emission centroids. This profile is then modeled according to the following equation:

\begin{equation}
R(\theta) = R_{0}\mathrm{csc}\theta\sqrt{3\left(1-\theta\mathrm{cot}\theta \right)},\
\end{equation}
\noindent
where $R_{0}$ is the stand-off distance between the apex of the bowshock and the position of the star and $\theta$ the different opening angles of the bow-shock profile. $R_{0}$ depends on the pressure-momenta balance between the relative motion of the stellar wind and the circumstellar medium as follows:

\begin{equation}
R_0 = \sqrt{\frac{\dot{m}v_w}{4\pi\rho v_*^2}},\
\end{equation}
\noindent
where $\dot{m}$ is the mass-loss rate of the star, $v_w$ is the terminal velocity of the wind, $\rho$ is the mean density of the circumstellar medium, and $v_*$ is the relative velocity of the star through the environment.

To determine $R_0$ we used a Monte-Carlo Markov-Chain model fitting using the \textit{emcee} code \citep{ForemanMackey13}, leaving $\dot{m}$ and the projected position angle of the bow shock in the plane of the sky as free parameters. $v_w$ was set to 2000 km s$^{-1}$, a typical parameter for  massive stars \citep[e.g.,][]{KP2000}. $\rho=\mu m_\mathrm{H} n_\mathrm{H_2}$ 
was obtained from previous mass estimations of the W49N molecular clump which give an average density $n_\mathrm{H_2} \sim 10^4$ cm$^{-3}$ \citep{GM13}, and using $\mu=2.4$ as the mean molecular weight of the medium. 
$v_*$ is set to 76 km s$^{-1}$ as determined from the observed proper motions here reported. It should be noted that most massive stars are part of binary systems and that the velocity adopted here is actually the velocity
of the bowshock. The true velocity of the star with respect to the medium could be modulated by motions around a companion.

\begin{figure}
\centering
\includegraphics[angle=-90,scale=0.6]{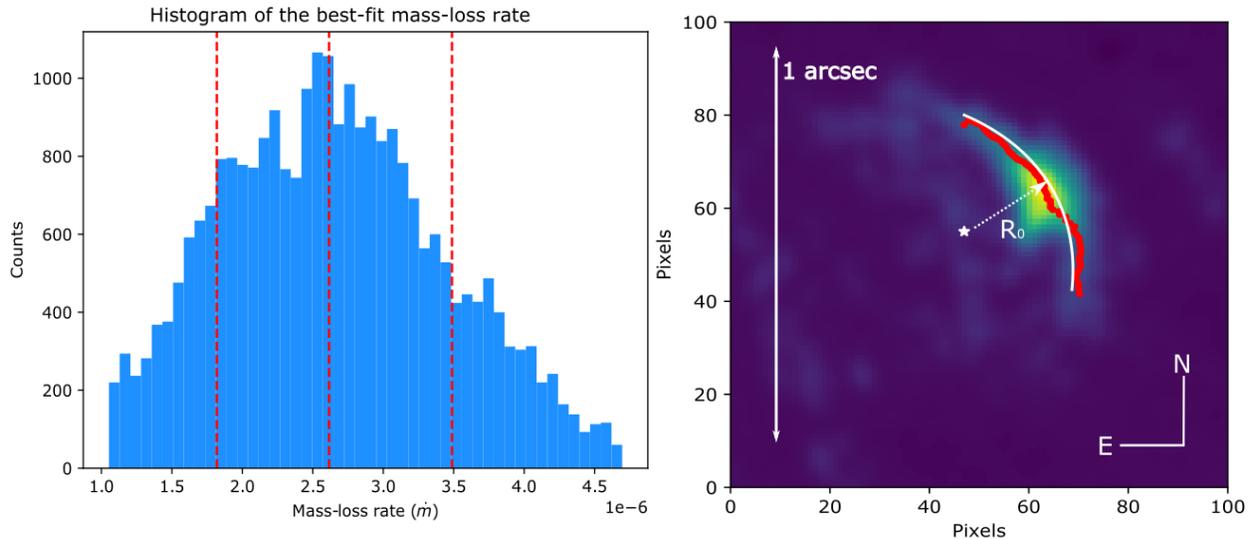}
\vskip-2.7cm
\caption{\small \textit{Left:} Histogram of the best-fit mass-loss rate. The vertical red-dotted lines indicate the 16\%, 50\% and 84\% of the percentiles in the distribution. The mass loss rate is given in units of $M_\odot~yr^{-1}$. \textit{Right:} Image of the bow shock with the extracted profile (red-line); the best-fit model of the shape  (white-line); the position of the central source (white star) and; the position of the bow-shock apex (dotted-white arrow). The angular scale of the image and its orientation are displayed in the panel.
}
\label{fig5}
\end{figure}

Figure \ref{fig5} displays the histogram of the best-fit values for the mass-loss rate. We obtain a mass-loss rate $\dot{m}$=2.62$\times$10$^{-6} \pm$8.7$\times$10$^{-7}$ M$_{\odot}$/yr, which corresponds to a stand-off distance of $R_0$=2216 $\pm$368 au at a P.A. of 300$^{\circ}$ (E$\rightarrow$N). 
Figure \ref{fig5} also displays the bowshock with the best-fit profile over-plotted on the radio continuum map, as well as the derived position of the central star.

We use the model-derived $\dot{m}$ to estimate 
that the runaway has a stellar mass $M_\star \approx 33\pm2~M_\odot$ \citep{Dzib13,Vacca96}, corresponding to a main-sequence spectral type of O6$\pm$0.5.
We can compare this spectral type estimate with that obtained by assuming that all the photoionization of component C is due to the UV radiation of the central star. The free-free flux density of component C at 1.3 cm is 0.45 Jy (DePree et al. 2000). Assuming that the emission is optically-thin and that the electron temperature is $10^4$ K, we derive that an ionizing photon rate of $5.8\times10^{48} ~s^{-1}$ is required to keep the photoionization \citep{Snell19}. An O8 ZAMS star can provide this photoionization \citep{Dzib13}. An explanation for the apparent discrepancy between the spectral types derived from the mass loss rate and the ionizing photon rate is that most of the ionizing photons of the central star escape to its SE, where there is no gas to stop the photons. If we assume that only 1/3 of the photons from the central star are absorbed by component C, an O6 ZAMS star is then required and excellent agreement is obtained between the two independent estimates.
The assumption that only 1/3 of the ionizing photons are intercepted by component C is consistent with the geometry shown in the right panel of Fig. \ref{fig5}. In this Figure the opening angle of component C as seen from the central star $\theta \simeq 120^\circ$. The fraction of the 4$\pi$ solid angle around the central star that is intercepted by component C is $(1/2)(1-cos(\theta/2) \simeq 0.25$, consistent with the required value of 0.3.


\citet{DePree04} report the local standard of rest (LSR) H$52\alpha$ radial velocity of 15 ultracompact HII regions in the zone. Excluding component C, the mean and rms values of these velocities are 11.9$\pm$6.5 km s$^{-1}$. Component C has
$v_{LSR}$ = -8.0$\pm$0.7 km s$^{-1}$, blueshifted by more than 3$\sigma$ away from the mean value. 
This confirms that component C has different motions than the rest of the stellar sources and the dense-gas clump and provides strong supporting evidence of the anomalous kinematics of component C. 

The potential well of the W49N region is given by the sum of the molecular gas mass within a 3 pc radius $M_\mathrm{gas} \approx 1\times10^5~M_\odot$ and the more uncertain embedded stellar mass $M_\star \sim 10^4~M_\odot$ \citep{GM13}. Thus, the escape velocity from the embedded cluster is $v_\mathrm{esc}\approx 17.8$ km s$^{-1}$. 
The 3D velocity of source C with respect to the stellar cluster is $v_\mathrm{C,3D} \approx 78.6$ km s$^{-1}$. Therefore, source C is clearly able to escape from the W49N clump region and constitutes a massive, runaway young star. If its velocity remains unchanged, source C could escape from the W49N clump within the next 4 to $5\times10^4$ yr, which is an order of magnitude smaller than the $\lesssim 1$ Myr age of massive embedded clusters such as W49N. 

It is possible that ejection events like the one we report are not uncommon during the span of star cluster formation. Massive runaways have been reported from more evolved (optically visible, ages $> 1$ Myr) clusters such as R136 in the LMC \citep{Lennon18} and Cygnus OB2 \citep{Comeron07}, with travel times $\sim 1$ Myr, consistent with ours being an earlier example. 

\section{Summary}


We analyze high-angular resolution 7-mm VLA data of the Welch ring of UC HIIs in W49N obtained in epochs 1998.28 and 2016.74. We find that source C has large proper motions of $1.41\pm0.11$ mas yr$^{-1}$ toward the NW (P.A.$=-51^\circ \pm 4 ^\circ$), equivalent to a velocity in the plane of the sky of $76\pm6$ km s$^{-1}$. Together with its line-of-sight velocity $\approx -20$ km s$^{-1}$ blueshifted from the mean cluster velocity as previously measured from recombination lines, we infer that source C is a young, massive runaway star, which was ejected from the Welch ring $\sim 6400$ yr ago. We estimate the spectral type of this runaway star from both the mass loss rate and the ionizing photon rate and find that both estimates are consistent with an O6 ZAMS type.

We also report on a small flux density increase for sources B1, B2, G2a, and G2c, and interpret these changes in the context of previous observations and models of dynamic massive star formation. 

\clearpage

\clearpage

\clearpage

\clearpage

\acknowledgements
We thank an anonymous referee for suggestions that improved the 
clarity of the paper. LFR acknowledges the financial support of
PAPIIT-UNAM and of CONACyT (M\'exico). 
RGM acknowledges support from UNAM-PAPIIT project IN104319. 
CGD acknowledges support from NSF RUI award 1615311.
This research has made use of the SIMBAD database, operated at CDS, Strasbourg, France.

\facility{VLA}

\software{AIPS \citep{vanMoorsel96}; CASA \citep{McMullin07}; \textit{emcee} \citep{ForemanMackey13}.}
 
\bibliographystyle{yahapj.bst}
\bibliography{references.bib}

\end{document}